# Impact of Boron doping to the tunneling magnetoresistance of Heusler alloy Co$_2$FeAl

Ali Habiboglu, Yash Chandak, Pravin Khanal, Bowei Zhou, Carter Eckel, Jacob Cutshall
Kennedy Warrilow, John O'Brien, John R. Schaibley, Brian J. Leroy and Wei-Gang Wang*

Department of Physics, University of Arizona, Tucson, AZ 85721, USA

Heusler alloys based magnetic tunnel junctions can potentially provide high magnetoresistance, small damping and fast switching. Here junctions with Co$_2$FeAl as a ferromagnetic electrode are fabricated by room temperature sputtering on Si/SiO$_2$ substrates. The doping of Boron in Co$_2$FeAl is found to have a large positive impact on the structural, magnetic and transport properties of the junctions, with a reduced interfacial roughness and substantial improved tunneling magnetoresistance. A two-level magnetoresistance is also observed in samples annealed at low temperature, which is believed to be related to the memristive effect of the tunnel barrier with impurities.

*wgwang@arizona.edu

Heusler alloys are a remarkably vast category of intermetallic materials with 2:1:1 (full Heusler) or 1:1:1 (half Heusler) [1–3], covering a wide range of interesting material systems from superconductors,[4] topological insulators,[5] Weyl semimetals[6] to fully spin-polarized half-metals.[7] The very high spin polarization of Heusler alloys, plus the symmetry filtering feature[8] of tunnel barriers like MgO that offer high transmission rate only to electrons with a certain symmetry (e.g. $\Delta_1$ band electrons), lead to a high tunneling magnetoresistance (TMR) of 2610% in $Co_2$(Mn,Fe)Si-MgO magnetic tunnel junctions (MTJs) at low temperature,[9] which is substantially higher than the 1144% of CoFeB-MgO junctions.[10] Due to the absence of one spin species at the Fermi level, at least in principle, the magnetic damping parameter can also be very small in half metallic Heusler alloys, which is of great importance for spintronics applications. In addition, ultrafast magnetization switching has recently been demonstrated in ferrimagnetic Heusler alloy $Mn_2Ru_xGa$,[11] where upon the heating of a laser pulse on the film the exchange scattering effect of the two inequivalent magnetic sublattices results in the reversal of magnetization in a few picoseconds. Similar ultrafast switching has already been achieved in MTJs with rare-earth compound CoFeGd,[12,13] despite the small TMR. Therefore, it is expected that the next generation Heusler-based MTJs with picosecond spin dynamics will have unique advantages in a wide range of applications where non-volatility, high on-off ratio and ultrafast operation are simultaneously desired.

Co-based alloys have been of noteworthy interest for using in MTJs, especially $Co_2FeAl$ (CFA)[14], where high TMR ratios have been successfully demonstrated[15,16]. In these MTJs with in-plane magnetic anisotropy, the $L2_1$ or B2 phases of CFA give rise to the high spin polarization that is required for the large TMR[17]. In all these reports, the epitaxial CFA(001)-MgO(001) structures were obtained through the seeding layers below the CFA, whose epitaxy was from the MgO(100) or Si(100) substrates. High temperature deposition or intermittent high temperature annealing before the completion of MTJ deposition were required. On the contrary, a simpler method is usually employed in the fabrication of CoFeB/MgO/CoFeB MTJs, [10,18] where the deposition of the entire MTJ stack can be done at room temperature (RT), without the complicated high temperature epitaxy from the substrate. In this method, the epitaxial CoFeB(001)/MgO(001)/ CoFeB(001) structure is achieved in the post-growth thermal annealing, in which the TMR increases from tens of percent to hundreds of percent through the solid state epitaxy process. [19–22]

Here we report the study on the CFA-MgO MTJs fabricated at RT on Si/SiO2 substrates. Boron doped CFA films are fabricated by co-sputtering. Similar to the case of CoFeB, the addition of Boron in the CFA has a profound impact on the structural, magnetic and transport properties of the MTJs. A reduced roughness is achieved in doped films, consistent with the exception of smooth interface of amorphous films due to the absence of lattice structure and grain boundaries. The dependence of TMR on the Boron doping concentration is investigated. A multiple level TMR effect is also observed in MTJs annealed at low temperatures, presumably related to the memristive nature of MgO barrier with Al impurities.

The samples in this study are grown on a thermally oxidized Silicon substrate by magnetron sputtering. Two types of MTJs are investigated. In the first type the CFA layer is on the bottom of MgO (BOT-CFA-MTJ) with the stack structure of Si/SiO2/Ta(7)/Ru(11)/Ta(6)/ $(Co_2FeAl)_{1-x}B_x$ (4-5)/MgO(1.5-3.5)/$Co_2Fe_6B_2$ (4-5)/Ta(7)/Ru(20), where all thicknesses are in nm. In the second type the CFA layer is on the top of MgO (TOP-CFA-MTJ) with the stack structure of Si/SiO$_2$(Si sub.)/Ta(7)/Ru(11)/Ta(6)/Co$_2$Fe$_6$B$_2$(4-5)/MgO(1.5-3.5)/(Co$_2$FeAl)$_{1-x}$B$_x$ (4-5)/Ta(7)/Ru(20). The base pressure of the system is on the order of 10$^{-9}$ Torr. Atomic force microscope (AFM) is used to compare the surface roughness of (Co$_2$FeAl)$_{90}$B$_{10}$ and Co$_2$FeAl. X-Ray Diffraction (XRD) measurement is done to characterize the crystal structure of the sample Si/SiO$_2$/Co$_2$FeAl(12). Metallic targets are sputtered by with DC sources and MgO is deposited by a rf source. The MTJs are patterned using conventional microfabrication techniques into elliptical shapes, then annealed in a rapid thermal annealing setup in Ar environment. During TMR measurements the magnetic field is applied along the long axis of the ellipse. More details of the sample fabrication can be found in our previous publications[23–28].

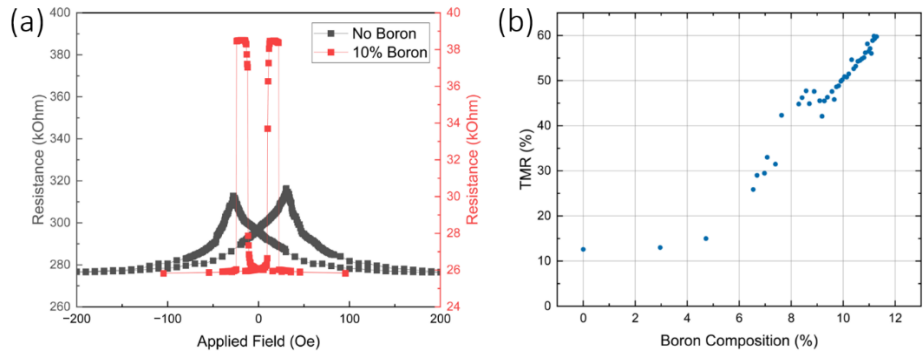

Figure 1 (a) Representative TMR curves of the BOT-CFA-MTJs with and without Boron doping. (b) TMR dependence on Boron composition (in percentage) in the BOT-CFB-MTJs.

The CFA layer is incorporated in the MTJs, initially as the bottom ferromagnetic (FM) electrode below MgO, replacing the bottom CoFeB layer in conventional CoFeB/MgO/CoFeB structures. These MTJs were annealed at 500 °C for 10 min prior to transport measurement. The BOT-CFA-MTJ exhibits a very low TMR about 12% as shown in Fig. 1(a). The low TMR can be due to the nearly diminished coherent tunneling effect as a result of a very bad CFA/MgO interface, or the absence of the antiparallel (AP) plateau that is more of a magnetic origin. Clearly the latter case is more possible for the undoped BOT-CFA-MTJ. Remarkably, the addition of Boron dramatically improves the TMR as shown in the same figure. Sharp magnetization switching occurs for both the (Co$_2$FeAl)$_{1-x}$B$_x$ layer and CoFeB layer, with distinct switching fields. As a result, the TMR is substantially larger than the undoped junctions. This is a general behavior in a large range of samples, as shown by the Boron concentration dependence of TMR in Fig. 1(b). It can be clearly seen that TMR monotonically increases with Boron doping, up to the highest doping concentration of this study (about 11.5%). It is interesting to note that the absence of Boron has a negative impact not only on the switching of the CFA layer below MgO, but also the switching of the CoFeB layer on top of MgO. The lack of sharp magnetic switching for both layers indicates the CFA and CoFeB may be magnetically coupled.

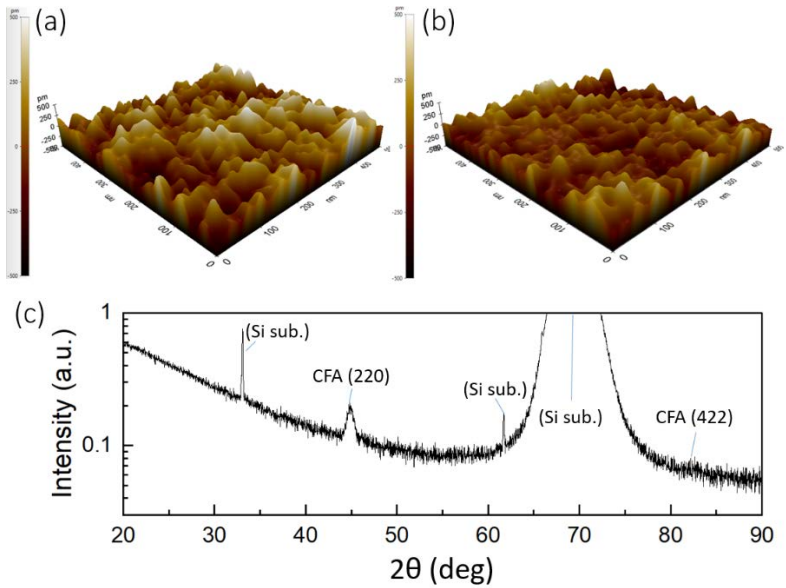

Figure 2: (a) AFM measurement of Buffer/Co$_2$FeAl and (b) Buffer/(Co$_2$FeAl)$_{90}$B$_{10}$ with roughness values 160pm and 120pm, respectively. (c) XRD measurement of Si/SiO2/Co$_2$FeAl(12nm)

It is known that orange peel coupling usually exists in thin films with large roughness.[29] To verify this, the roughness of the undoped and doped CFA films are measured by AFM in an oxygen free environment. Two samples are fabricated for this purpose, with the stack structure of Si sub./Ta(7)/Ru(11)/Ta(6)/Co$_2$FeAl(15) and Si sub./Ta(7)/Ru(11)/Ta(6)/(Co$_2$FeAl)$_{90}$B$_{10}$ (15). The roughness of the undoped CFA is found as 0.16nm as shown by Fig. 2(a). On the other hand, the roughness of CFA with 10% boron composition is found to be substantially smaller at 0.12nm as shown in Fig. 2(b). This 0.04nm difference in roughness, though not very big in magnitude, may have a large impact on the magnetic and transport properties of the MTJs. Therefore, the low TMR in the undoped MTJ may be at least partly due to the roughness induced orange peel coupling that leads to the absence of the AP state. In addition, it is well known that even in MTJs with well separated soft and hard layers, the TMR would be reduced by increased roughness, for example, in samples subjected to prolonged annealing. [22] An X-ray diffraction study is further carried out to investigate the crystalline structure of the CFA film, with an 12nm CFA film deposited on Si/SiO$_2$ substrate. As shown in Fig. 2(c), both the (220) and (422) diffraction peaks of CFA can be identified. Unfortunately, the (001) orientation that is needed for symmetry filtering tunneling with the MgO barrier, is not observed. Compared to the (Co$_2$FeAl)$_{90}$B$_{10}$ that is amorphous, the crystalline structure of CFA is one of the reasons for its larger roughness.

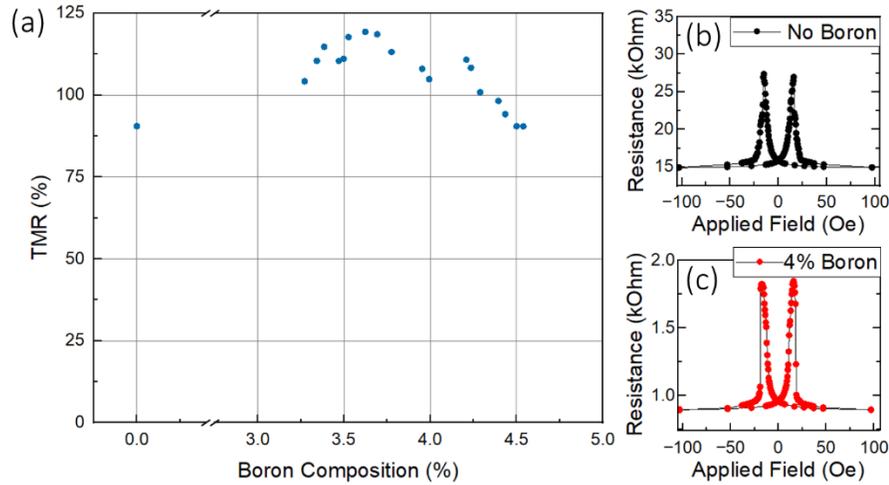

Figure 3: TMR measurements are collected from Top CFA samples (Buffer/Co$_2$Fe$_6$B$_2$(5)/MgO(2)/ (Co$_2$FeAl)$_{(1-x)}$B$_{(x)}$(4.8)/Capping) as the Boron composition increases (a). Two TMR curves for 2 different Boron compositions are shown as (b) x=0 and (c) x=3.7.

Next the TOP-CFA-MTJs are investigated. The Boron concentration is varied from 0 to 4.5% in the top CFA layer, while the bottom FM layer is CoFeB. The behavior of the TMR is plotted in Fig. 3(a). A number of interesting observations can be made immediately. Similar to the BOT-CFA-MTJs, Boron doping generally enhances the TMR. However, unlike the almost monotonic dependence of TMR on Boron concentration in the BOT-CFA-MTJs, the TMR peaks at about 3.5% of Boron in the TOP-CFA-MTJs. Moreover, the TMR ratios are generally higher in the TOP-CFA-MTJs, with 89% in the undoped junction and 120% in the junction with 3.7% of Boron, compared to the 13-15% TMR in the BOT-CFA-MTJs. The large difference in the TMR ratios in the two types of samples highlight the more critical function of the bottom FM electrode in the MTJs. The optimized CoFeB/MgO bottom interface is presumably smoother, [30,31] supporting a better MgO barrier and therefore able to provide a higher TMR ratio compared to the (Co$_2$FeAl)$_{1-x}$B$_x$/MgO bottom interface. The highest TMR of 120% in this study is substantially smaller than the 360% TMR obtained in a previous study with CFA/MgO/CFA MTJs[16]. The exact reason is not clear at this point. Note in Figure 3(c), there is no clear AP plateau observed. Even though the CFA-MTJ fabricated in this study by solid state epitaxy can provide high spin polarization as the junctions previous fabricated by high temperature deposition, the lack of AP will skew the real TMR value from the coherent tunneling. Exchange-bias by IrMn or PtMn will be employed in our forthcoming study to stabilize the AP state. Further study on the microstructure of the MTJ will also be very helpful.

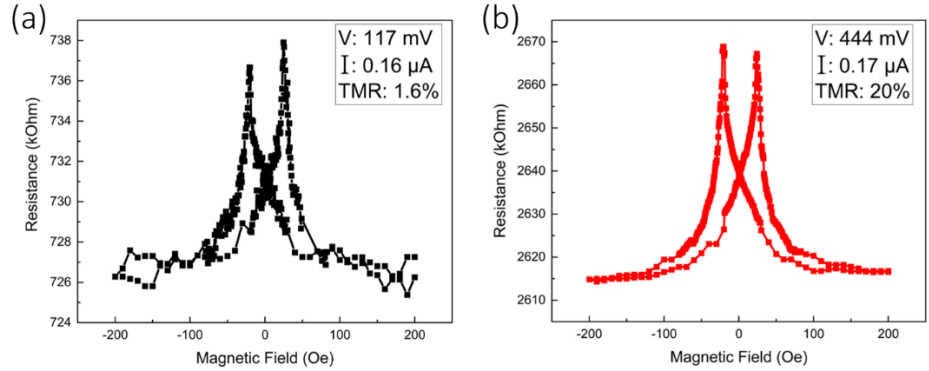

Figure 4: Multilevel TMR effect in an undoped BOT-CFA-MTJ annealed at 300 °C, where the low parallel resistance state with 1.6% TMR is measured at 117 mV (a), and the high parallel resistance state with 20% TMR is measured at 444 mV (b).

Finally, an interesting two-level TMR effect is observed in the undoped BOT-CFA-MTJs after annealing at 300 °C, as shown in Figure 4. A 20% TMR with smaller noise is observed with a higher parallel resistance at about 2.6 MΩ, compared to a 1.6% TMR at a lower parallel resistance at 726 kΩ. The high R state was achieved at the voltage of ~400 mV while the low R state was achieved at about 100 mV. Typically, the TMR ratio has a bias dependence in which TMR drops with increasing bias voltages. This is because of excitation of magnon[32] or impurity [33,34] that flips the spin of tunneling electrons, therefore the TMR is reduced at high voltages. Another reason for the bias dependence of the TMR is the dependence of the density of state (DOS) on voltage.[35] Here the case is opposite, a larger TMR is obtained at 444 mV instead of 117 mV. This is likely due to the improvement of the stoichiometry of the MgO barrier at the higher voltage. It is known that Al in the CFA barrier can diffuse into the MgO barrier, resulting in a non-stoichiometric $MgAlO_x$ barrier[36]. Therefore, it is possible that oxygen vacancies move under 400 mV, leading to a barrier with less spin-flip scattering caused by impurities, thereby giving rise to a larger effective spin-polarization. Even though larger number of magnons are excited at 400 mV compared to at 100 mV, the enhanced effective spin-polarization still dominates, therefore a large TMR was observed. Indeed, this effect is not observed if the BOT-CFA-MTJs are annealed at 500 °C, where the impurities/defects in the barrier are reduced with further annealing.[37] This is consistent with previous findings that metallic Al impurities in the barrier decreases as the annealing temperature increases due to the formation of $MgAlO_x/AlO_x$ barrier[36].

In summary, we have successfully prepared CFA-based MTJs on $Si/SiO_2$ substrates by room temperature sputtering. The doping of Boron in CFA has a strong influence on the TMR, related to the decrease in roughness of the $(Co_2FeAl)_{1-x}B_x$ films. In the TOP-CFA-MTJs, the TMR value is limited by the absence of clear AP states. Additionally, a memristive like behavior is observed in the MTJs that are annealed at low temperature. These results demonstrate a new way to achieve advanced MTJs based on Heusler alloys.

**Acknowledgments**

This work was supported in part by DARPA through the ERI program (FRANC), and by NSF through DMR-1905783. J.R.S. and B.J.L. were supported by the Army Research Office under Grant no. W911NF-18-1-0420. C. E. and K.W. were supported by the REU supplement of NSF ECCS-1554011.

**Data availability**

The data that support the findings of this study are available from the corresponding author upon reasonable request.